\documentclass[aps,prf,reprint,amsmath,amssymb,onecolumn]{revtex4-2}
\usepackage{graphicx}
\usepackage{tikz}
\usetikzlibrary{positioning, arrows.meta, shapes.geometric}
\usepackage{subcaption}
\usepackage{soul}
\usepackage{listings}
\usepackage{xcolor}
\usepackage{booktabs}
\usepackage{multirow}
\usepackage{array}
\usepackage{float}
\usepackage{makecell}
\usepackage{footnote}
\usepackage{CJKutf8}
\usepackage{algorithm2e}
\usepackage[margin=2.5cm]{geometry}
\usepackage{siunitx}

\RestyleAlgo{ruled}
\SetAlgoLined
\SetKwComment{Comment}{\color{blue}//}{}
\SetKwInput{KwInput}{\textbf{Input}}
\SetKwInput{KwOutput}{\textbf{Output}}

\begin{document}

\title{FeaGPT: an End-to-End agentic-AI for Finite Element Analysis}

\author{Yupeng Qi (\begin{CJK*}{UTF8}{gbsn}亓宇鹏\end{CJK*})}
\affiliation{
Cluster of Excellence SimTech, University of Stuttgart, Stuttgart, Germany
}

\author{Ran Xu (\begin{CJK*}{UTF8}{gbsn}徐冉\end{CJK*})}
\affiliation{
Faculty for Aerospace Engineering and Geodesy, University of Stuttgart, Stuttgart, Germany
}

\author{Xu Chu (\begin{CJK*}{UTF8}{gbsn}初旭\end{CJK*})}
\email{xu.chu@exeter.ac.uk}
\thanks{Corresponding author}
\affiliation{
Faculty of Environment, Science and Economy, University of Exeter, Exeter EX4 4QF, United Kingdom
}
\affiliation{
University of Stuttgart, Stuttgart, Germany
}

\date{\today}

\begin{abstract}
Large language models (LLMs) are establishing new paradigms for engineering applications by enabling natural language control of complex computational workflows. This paper introduces FeaGPT, the first framework to achieve complete geometry-mesh-simulation workflows through conversational interfaces. Unlike existing tools that automate individual FEA components, FeaGPT implements a fully integrated Geometry-Mesh-Simulation-Analysis (GMSA) pipeline that transforms engineering specifications into validated computational results without manual intervention. The system interprets engineering intent, automatically generates physics-aware adaptive meshes, configures complete FEA simulations with proper boundary condition inference, and performs multi-objective analysis through closed-loop iteration.

Experimental validation confirms complete end-to-end automation capability. Industrial turbocharger cases (7-blade compressor and 12-blade turbine at \SI{110000}{rpm}) demonstrate the system successfully transforms natural language specifications into validated CalculiX simulations, producing physically realistic results for rotating machinery analysis. Additional validation through 432 NACA airfoil configurations confirms scalability for parametric design exploration. These results demonstrate that natural language interfaces can effectively democratize access to advanced computational engineering tools while preserving analytical rigor.
\end{abstract}

\maketitle

\section{Introduction}

The emergence of Large Language Models (LLMs) represents a new paradigm in artificial intelligence for addressing computational problems across scientific and engineering disciplines. From early transformer architectures \cite{vaswani2017attention} to contemporary models including GPT-4 \cite{openai2023gpt4}, Claude \cite{anthropic2024claude}, and LLaMA \cite{touvron2023llama}, LLMs have demonstrated capabilities in understanding technical contexts, reasoning about physical systems, and generating solutions. These developments have enabled advances in fields ranging from drug discovery to materials science, from circuit design to computational mechanics.

The impact of LLMs on scientific computing has been particularly profound. In drug discovery, recent systems such as ChemCrow integrate large language models with specialized chemistry toolkits to accelerate molecular design workflows \cite{brown2020language,bran2023chemcrow}. Materials science has witnessed the development of specialized models for molecular representation and property prediction, with the 2024 LLM Hackathon for Materials Science producing 34 successful implementations across seven application domains \cite{lewis2020retrieval}. Multi-agent LLM systems have emerged as powerful tools for autonomous scientific discovery: SciAgents demonstrates automated materials design through bioinspired graph reasoning \cite{ghafarollahi2025sciagents}, while MechAgents successfully solves mechanics problems through collaborative agent interactions \cite{ni2024mechagents}, and AtomAgents achieves physics-aware alloy design with enhanced properties \cite{ghafarollahi2024atomagents}. In computational biology, models like HyenaDNA have pushed the boundaries by processing genomic sequences up to 1 million tokens, while specialized transcriptomic LLMs including Geneformer and scGPT have demonstrated remarkable abilities in analyzing complex biological data \cite{poli2023hyenadna,theodoris2023geneformer,cui2024scgpt}. Together with advances such as physics-informed neural networks for scientific computing \cite{raissi2019physics} and PDE solver generation frameworks \cite{li2025codepde}, these successes demonstrate the potential of LLMs to facilitate scientific discovery and improve research efficiency.

Engineering applications have similarly experienced advances through LLM integration. In electronic design automation, systems like LaMAGIC leverage language models for analog circuit topology generation, while domain-adapted LLMs such as ChipNeMo enable efficient chip design workflows \cite{liu2023chipnemo}. The CAD domain has witnessed transformative advances: Text2CAD generates sequential parametric CAD models from natural language prompts spanning beginner to expert levels \cite{khan2024text2cad}, while multimodal frameworks like LLM4CAD enable 3D design generation from combined text-image inputs \cite{li2025llm4cad}, and unified systems such as CAD-MLLM integrate diverse input modalities including point clouds for comprehensive CAD automation \cite{xu2024cadmllm}. Earlier pioneering work established foundational capabilities for zero-shot text-to-shape generation \cite{sanghi2022clipforge,poole2022dreamfusion}. These developments collectively demonstrate the potential of LLMs to democratize access to sophisticated engineering tools.

Within computational mechanics specifically, the integration of LLMs has shown exceptional promise. OpenFOAMGPT \cite{pandey2025openfoamgpt, wang-2025} established a successful paradigm for RAG-augmented LLM agents in computational fluid dynamics, achieving robust convergence across diverse scenarios including multiphase flows, heat transfer, and turbulence modeling. OpenFOAMGPT 2.0 \cite{feng2025openfoamgpt2} advanced this framework with a four-agent collaborative architecture enabling conversation-driven simulation workflows, achieving 100\% success rate in fully automated end-to-end CFD execution. Building on these foundations, turbulence.ai \cite{feng2025turbulenceai} introduced the first fully autonomous AI scientist for fluid mechanics, unifying hypothesis generation, numerical experimentation, and manuscript drafting into a single multi-agent system capable of producing publication-level scientific insights from natural language queries. Domain-specific fine-tuning has proven highly effective: AutoCFD employs a custom NL2FOAM dataset of 28,716 natural language-to-OpenFOAM configuration pairs with chain-of-thought annotations to enable direct translation from descriptions to executable CFD setups \cite{dong2025autocfd}. Multi-agent architectures have emerged as a dominant paradigm, with MetaOpenFOAM integrating RAG-enhanced knowledge retrieval and role-specialized agents to decompose complex tasks \cite{chen2024metaopenfoam}, ChatCFD implementing iterative trial-reflection-refinement mechanisms for handling complex literature-derived cases \cite{fan2025chatcfd}, and Foam-Agent employing hierarchical multi-index RAG with dependency-aware generation to ensure configuration consistency across the complete workflow \cite{yue2025foamagent}. Systematic evaluations reveal that LLMs exhibit strong capabilities in leveraging and applying domain-specific CFD knowledge to solve computational problems \cite{wang2025evaluations}. These systems have proven that LLMs, when integrated with domain-specific knowledge and computational tools, can successfully navigate the intricate requirements of physics-based simulations.

Despite these advances, structural finite element analysis (FEA) presents unique challenges that existing systems have not fully addressed. FEA constitutes the cornerstone of modern structural engineering, enabling prediction of mechanical behavior under complex loading conditions \cite{zienkiewicz2005finite}. The complexity of FEA workflows—spanning geometry creation, mesh generation, boundary condition specification, solver configuration, and result interpretation—requires expertise across multiple domains \cite{hughes2012finite}, which presents accessibility challenges for broader engineering communities. AutoFEA \cite{hou2025autofea} represents an important advance by combining GCN-Transformer models with LLMs for automated code generation, demonstrating the viability of LLM-assisted FEA workflows. However, complete automation from natural language to validated designs—encompassing geometry generation, mesh creation, solver configuration, and result interpretation—represents the natural next step toward truly democratized FEA access. While specific workflow components have seen advances (DeepFEA for solver acceleration, learning-based mesh generation), the integration of these capabilities into a unified natural language interface would enable rapid design iteration and establish a new paradigm for intelligent computer-aided engineering tools.

This paper presents FeaGPT, a natural language-driven framework for structural finite element analysis. The system automates the complete workflow from geometry creation to simulation results. Users provide textual descriptions of structures, loading conditions, and material properties. The framework generates geometries intelligently—leveraging validated knowledge for standard components while synthesizing novel designs as needed—creates finite element meshes, configures solver parameters, executes simulations, and extracts results. The system includes an analysis module that can process parametric study results from batch simulations to identify optimal design configurations.

\section{Methods}

In this section, we present the technical details of FeaGPT, our proposed framework for automating finite element analysis through natural language interfaces. The overall architecture is designed to transform free-form engineering descriptions into validated FEA results without manual intervention.

\subsection{System Overview}

FeaGPT implements a modular architecture organized around the GMSA (Geometry-Mesh-Simulation-Analysis) pipeline. This sequential structure reflects the natural FEA workflow, extended with data-driven analysis for extracting insights from simulation results. Each module operates independently while maintaining data consistency through standardized interfaces.

The system transforms natural language input into validated engineering results through five integrated stages:

\begin{enumerate}
\item \textbf{Engineering Analysis Planning}: Analyzes engineering intent, formulates comprehensive simulation strategies, and orchestrates computational workflows from textual descriptions
\item \textbf{Geometry Generation}: Converts simulation requirements into parametric CAD models using FreeCAD's Python API
\item \textbf{Intelligent Meshing}: Applies physics-aware adaptive mesh generation strategies through Gmsh
\item \textbf{FEA Simulation}: Configures and executes CalculiX solver for structural simulation, producing stress fields, displacements, and other engineering metrics
\item \textbf{Result Analysis}: For parametric studies, applies appropriate data-driven methods based on user requirements—including regression analysis, clustering, sensitivity analysis, surrogate modeling, or multi-objective optimization \cite{breiman2001random,rasmussen2006gaussian,saltelli2008global,deb2002fast,marler2004survey}—to extract actionable insights from simulation results
\end{enumerate}

The specific workflow depends on task complexity: single-configuration analyses conclude after simulation (step 4) with direct FEA results, while parametric studies or design optimization tasks proceed through result analysis (step 5). The analysis method is determined by natural language input—for instance, "find the optimal design" activates multi-objective optimization, "understand parameter sensitivity" invokes sensitivity analysis, or "identify failure patterns" employs clustering algorithms. This integrated pipeline enables the system to handle tasks ranging from simple structural verification to comprehensive design space exploration.

\begin{figure}[htbp]
\centering
\includegraphics[width=0.95\textwidth]{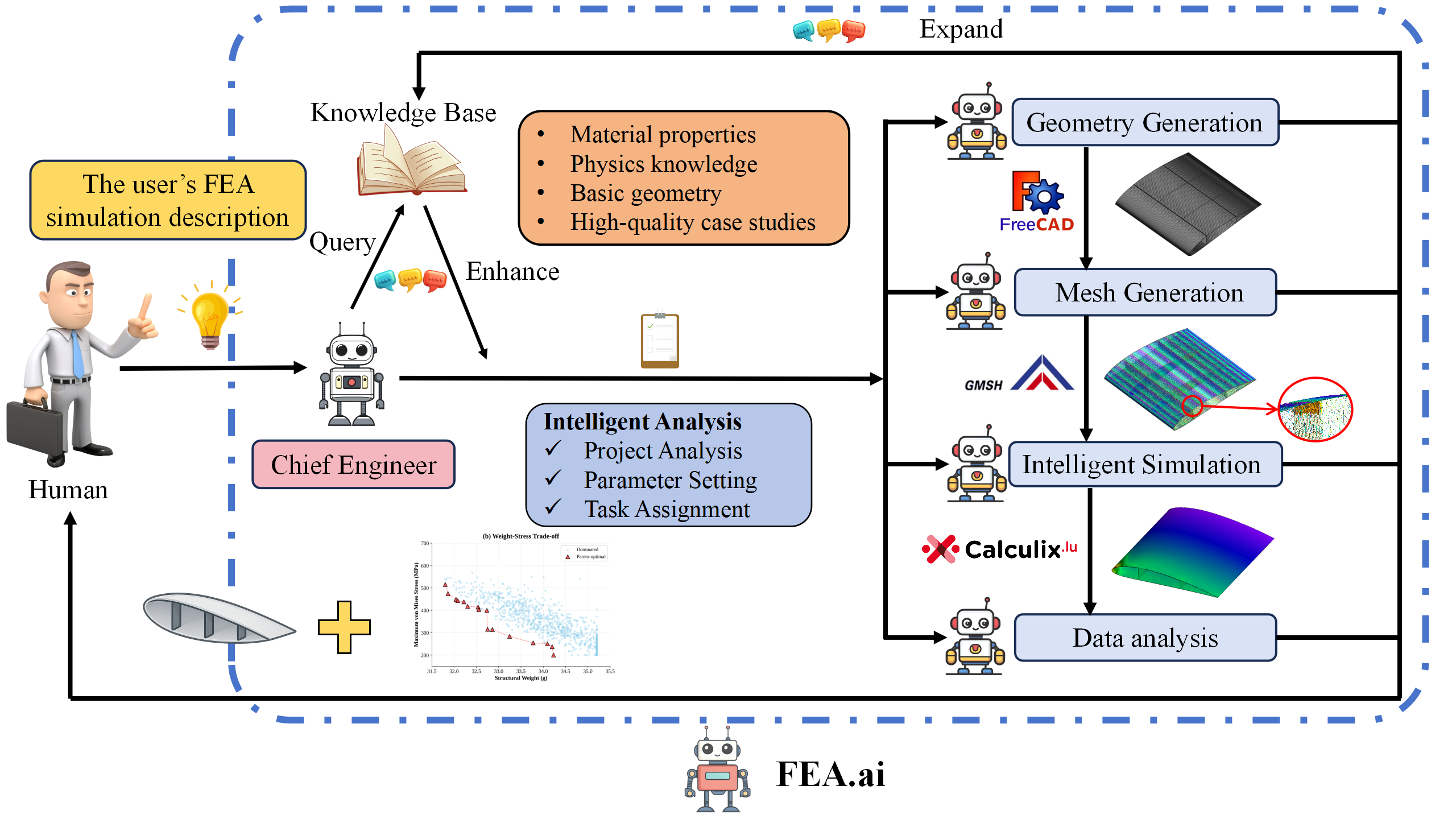}
\caption{FeaGPT system architecture illustrating the complete GMSA (Geometry-Mesh-Simulation-Analysis) pipeline. The framework transforms natural language specifications into validated FEA results through intelligent analysis planning, adaptive geometry generation, physics-aware meshing, and solver configuration, with optional data-driven analysis for parametric studies.}
\label{fig:system_architecture}
\end{figure}

Figure~\ref{fig:system_architecture} illustrates the complete FeaGPT architecture, showing how natural language input flows through the system to produce validated FEA results. The framework integrates a comprehensive engineering knowledge base encompassing validated geometric patterns, material properties, simulation strategies, and solver configurations accumulated from successful prior analyses. When generating geometry, the system leverages this knowledge base ($\sigma > 0.85$ similarity threshold) for standard components while synthesizing novel designs as needed, balancing reliability with flexibility across diverse engineering scenarios. The framework employs a vector-indexed knowledge base for efficient semantic search. For parametric studies, all configurations retrieve from this unified knowledge source, ensuring consistent access to validated geometric definitions, material properties, and solver configurations.

\subsection{Engineering Analysis Planning and Task Orchestration}

The natural language layer functions as an intelligent project manager that analyzes engineering intent, formulates comprehensive analysis plans, and orchestrates computational tasks. The system performs three critical functions: (1) \textit{Engineering Context Analysis}—interprets implicit constraints and domain conventions (e.g., "aerospace application" implies material selection, loading patterns, and safety requirements); (2) \textit{Analysis Strategy Formulation}—develops comprehensive analysis plans (e.g., "fatigue assessment" triggers multi-step load spectrum with S-N curve evaluation); (3) \textit{Task Orchestration}—decomposes complex requests into executable computational tasks with proper dependency management. For example, "analyze a NACA0012 wing with 3 spars under aerodynamic loads" is analyzed to determine: airfoil geometry specification, internal structural configuration, load distribution strategy, constraint patterns, and analysis sequence—all autonomously planned by the system.

The system implements knowledge-augmented LLM for analysis planning and task orchestration. This process transforms natural language requirements into structured task allocation interfaces: material specifications are allocated to the material configuration module, loads and boundary conditions to the boundary condition generation module, and analysis directives to the solver configuration module. The key innovation lies in semantic topology mapping: rather than extracting precise coordinates, loads and boundary conditions are specified through semantic location descriptors ("left edge", "hole boundary"), which the geometry analyzer subsequently maps to actual surface patches. Material properties are retrieved dynamically through vector search: based on contextual understanding, the system queries the knowledge base in real-time (e.g., "aerospace aluminum" automatically retrieves complete mechanical properties of Al 7075-T6). Algorithm~\ref{alg:analysis_planning} illustrates this planning process and its structured output format.

\textbf{Engineering Requirements Analysis:} The system analyzes engineering requirements through multi-level interpretation:
\begin{itemize}
\item \textit{Quantitative specifications}: Dimensional values with unit recognition and physical consistency checking (chord length, span, thickness with mm/m/in conversion)
\item \textit{Design space definition}: Range specifications automatically expanded to complete parameter spaces for comprehensive exploration (e.g., "1.0-2.0 mm, step 0.5" generates parametric study configurations)
\item \textit{Performance objectives}: Goal statements translated to computational metrics (e.g., "lightweight" $\rightarrow$ weight minimization, "high stiffness" $\rightarrow$ displacement constraints)
\item \textit{Analysis requirements}: Problem type inference from engineering context (static, modal, fatigue, buckling)
\end{itemize}

\textbf{Component Pattern Recognition:} The system identifies engineering components through pattern analysis and semantic understanding, enabling intelligent geometry planning. For instance, detecting "NACA0012" versus "NACA4412" triggers appropriate airfoil profile formulations with corresponding geometric definitions.

\textbf{Analysis Strategy Formulation:} For comprehensive analysis planning, the system employs Gemini 2.5 Pro to formulate complete analysis strategies from natural language descriptions. The LLM comprehends engineering semantics and generates structured analysis plans containing:

\begin{itemize}
\item \textit{Material specification}: Name recognition with automatic database matching and property retrieval (e.g., "Aluminum 7075-T6" $\rightarrow$ E=71.7 GPa, $\nu$=0.33, $\rho$=2810 kg/m³, $\sigma_y$=503 MPa)
\item \textit{Loading strategy formulation}: Magnitude, distribution patterns, and semantic location planning (e.g., "500N on right edge" $\rightarrow$ \{type: force, magnitude: 500, direction: X, location: "right\_edge"\})
\item \textit{Constraint planning}: Boundary condition strategy based on structural function (e.g., "fix left edge" $\rightarrow$ \{type: fixed, location: "left\_edge", constraints: [X,Y,Z]\})
\item \textit{Analysis sequence determination}: Workflow planning from engineering context (static $\rightarrow$ modal $\rightarrow$ fatigue based on requirements)
\end{itemize}

This intelligent planning enables the system to infer complete analysis workflows from high-level descriptions, automatically determining required analysis sequences, loading conditions, and evaluation criteria based on structural type and engineering context. The system balances computational efficiency (pattern matching: $<$1ms) with semantic flexibility (LLM planning: ~1.5s, \$0.006/query).

\begin{algorithm}[h]
\caption{Analysis Planning and Task Orchestration via Structured JSON}
\label{alg:analysis_planning}
\KwInput{Natural language description $D$, Material knowledge base $\mathcal{K}_M$}
\KwOutput{Structured FEA specification $S$}

\textbf{Step 1:} Vector retrieval: $M \leftarrow \text{TopK}(\text{similarity}(D, \mathcal{K}_M), k=5)$\;
\textbf{Step 2:} Construct augmented prompt: $P \leftarrow \text{Template}(D, M)$\;
\textbf{Step 3:} LLM planning: $S \leftarrow \text{Gemini 2.5 Pro}(P)$\;

\textbf{Output Structure:}\\
\begin{lstlisting}[basicstyle=\ttfamily\footnotesize]
{
  "material": {
    "name": "Al 7075-T6",
    "youngs_modulus": 71.7e9,  // Pa
    "poissons_ratio": 0.33,
    "density": 2810,  // kg/m^3
    "yield_strength": 503e6  // Pa
  },
  "loads": [{
    "type": "force",
    "magnitude": 500,  // N
    "direction": "X",
    "location": "right edge"  // Semantic location
  }],
  "boundary_conditions": [{
    "type": "fixed",
    "location": "left edge",  // Semantic location
    "constraints": ["X", "Y", "Z"]
  }],
  "mesh": {
    "density": "fine",  // ultra_fine/fine/medium/coarse
    "element_type": "C3D10",  // 10-node tetrahedral
    "refinement_zones": ["hole boundary"]
  },
  "analysis": {
    "type": "static",
    "solver": "CalculiX"
  },
  "data_analysis": {
    "objectives": ["minimize stress", "minimize weight"],
    "metrics": ["von_mises_stress", "displacement", "mass"],
    "optimization": "pareto_front"  // single/parametric/pareto
  }
}
\end{lstlisting}
\end{algorithm}

The knowledge base contains accumulated engineering experience from verified workflows across three categories: material properties, validated analysis configurations, and error diagnosis patterns. Each successfully completed simulation contributes its specifications back to the knowledge base—geometric parametrization, material selection, boundary condition strategies, solver settings, and encountered error solutions—creating a self-improving system where accumulated experience continuously expands capability while maintaining reliability through validation.

The framework employs vector-based semantic search for knowledge retrieval. The system computes similarity $\sigma$ between analysis requirements $\mathcal{R}$ and knowledge base entries $\mathcal{K}$ using cosine distance in embedding space:

\begin{equation}
\sigma(\mathcal{R}, \mathcal{K}) = \max_{k \in \mathcal{K}} \cos(\text{embed}(\mathcal{R}), \text{embed}(k))
\end{equation}

The embedding function uses sentence-transformers (all-MiniLM-L6-v2) to generate 384-dimensional vectors from textual descriptions, enabling fuzzy semantic matching across multilingual terminology and diverse engineering contexts. When $\sigma > 0.85$, the system retrieves validated patterns from prior analyses, ensuring consistency with proven design practices. For novel configurations below this threshold, the system synthesizes new solutions while maintaining physical validity constraints.

Following parameter planning, the system executes geometry generation and FEA configuration in parallel. The JSON specification output from Algorithm~\ref{alg:analysis_planning} is allocated to material configuration, load application, and boundary condition modules, while geometry generation independently analyzes the original natural language description to determine geometric strategy. These parallel workflows converge at the final solver configuration stage, ensuring consistent mapping between geometric topology and analysis parameters.

\subsection{Intelligent Geometry Generation with Knowledge Augmentation}

FeaGPT generates complete geometric models directly from textual descriptions through intelligent synthesis. The system autonomously determines the optimal generation strategy based on engineering context—leveraging validated knowledge from successful prior analyses when applicable, while generating novel geometries when required. Rather than a rigid template-versus-generation dichotomy, the framework implements an intelligent continuum: the system references accumulated engineering knowledge (successful prior analyses) to ensure consistency with established practices, while maintaining the flexibility to synthesize novel designs. This knowledge-augmented approach ensures both reliability for standard analyses and innovation capability for unique problems.

\textbf{Intelligent Generation Strategy:} The system selects generation strategy based on semantic similarity between the analysis requirements and the knowledge base:

\begin{equation}
\text{Strategy} = \begin{cases}
\text{Knowledge-Augmented}, & \text{if } \sigma(\mathcal{R}, \mathcal{K}_T) > \theta \\
\text{Novel Synthesis}, & \text{otherwise}
\end{cases}
\end{equation}

where $\sigma$ measures the similarity score between the analysis plan $\mathcal{R}$ and knowledge base $\mathcal{K}_T$, and $\theta$ is empirically set to 0.85. When analyzing standard components ($\sigma > 0.85$), the system references validated geometric definitions from successful prior analyses—ensuring consistency with engineering standards (e.g., NACA airfoil specifications, ASME pressure vessel codes) while enabling reproducibility across parametric studies. For novel designs, the system synthesizes geometries through LLM-based code generation.

\textbf{Knowledge-Augmented Generation:} The system maintains a self-growing knowledge base of validated geometric patterns. Currently, this includes parametric generators for standard engineering components (such as airfoil profiles and structural elements), with automatic accumulation of new geometric definitions through each successful analysis. The knowledge base employs modular organization, supporting flexible composition from fundamental geometric primitives to complex assemblies. This extensible architecture enables the system to progressively cover broader engineering application scenarios.

The knowledge base encapsulates domain expertise validated through successful analyses. When applicable, the system references these proven definitions with parameterized generation. The system maintains mathematical accuracy by implementing exact formulations from engineering standards. For example, NACA 4-digit airfoil profiles implement the standard thickness distribution:

\begin{equation}
y_t = \frac{t}{0.2} \left[ 0.2969\sqrt{x} - 0.1260x - 0.3516x^2 + 0.2843x^3 - 0.1015x^4 \right]
\end{equation}

where $t$ is the maximum thickness ratio and $x$ is the normalized chord position \cite{abbott1959theory}. Similarly, turbine blade profiles incorporate proper twist and taper distributions, while pressure vessels implement thickness calculations based on ASME standards. The knowledge base includes intelligent structural element placement algorithms that automatically handle component interactions, generating appropriate cutouts where structural members intersect.

\textbf{Novel Synthesis Mode:} For geometries not matching established patterns, the system employs Gemini 2.5 Pro to generate FreeCAD Python scripts from natural language descriptions. The LLM receives structured prompts containing: (1) analysis requirements and constraints, (2) FreeCAD API documentation excerpts, and (3) four complete example scripts demonstrating common operations (boxes, cylinders, boolean cuts). Generated code undergoes three-layer validation: syntax checking via Python AST parsing, security validation through blacklist/whitelist pattern matching (18 forbidden operations including os.system, eval; 3 required patterns including FreeCAD import and STEP export), and geometric feasibility verification through FreeCAD's topology checkers. Successfully demonstrated on a mounting plate with center hole (150×100×8mm with Ø30mm hole), achieving 100\% success rate with proper boolean operation handling and volume validation (114,345 mm³, 0.00\% error). Generation time: 3-15 seconds including validation.

Generated scripts are validated for syntactic correctness and geometric feasibility before execution. Successfully validated novel geometries can be added to the knowledge base, enabling continuous knowledge accumulation.

\textbf{Technical Implementation:} The system generates FreeCAD Python scripts following a consistent structure. When referencing the knowledge base, the system performs parameter substitution on validated geometric definitions—for instance, adapting standardized formulations with user-specified values. For novel synthesis, the LLM receives structured prompts containing: (1) analysis requirements and constraints, (2) FreeCAD API documentation excerpts, and (3) example code patterns for common operations. Generated scripts undergo syntax validation via Python AST parsing, followed by geometric validation through FreeCAD's built-in collision and topology checkers. Both approaches output standardized STEP files for mesh generation, ensuring compatibility with downstream tools.

This intelligent approach achieves complementary strengths: knowledge-augmented generation provides fast, reliable creation for standard components (typical execution: 1-2 seconds), while novel synthesis handles arbitrary designs (typical execution: 8-15 seconds including generation and validation). The system seamlessly selects the appropriate strategy based on engineering context, ensuring both reliability for established analyses and innovation capability for novel problems.

\subsection{Adaptive Mesh Generation}

The mesh generation module receives the STEP geometry file and semantic mapping from the geometry analyzer. This semantic mapping establishes correspondence between semantic locations defined in Algorithm~\ref{alg:analysis_planning} (``left edge'', ``hole boundary'') and actual geometric surface IDs, ensuring consistent matching between mesh physical groups and load/boundary conditions. Additionally, the system extracts mesh density specifications (ultra\_fine/fine/medium/coarse) from the original natural language description and automatically configures corresponding mesh parameters as defined in the structured output.

Mesh quality directly impacts both solution accuracy and computational efficiency. Traditional approaches require manual specification of mesh parameters—a process that demands significant expertise and iterative refinement. FeaGPT automates this through intelligent mesh adaptation using Gmsh \cite{geuzaine2009gmsh}, which automatically adjusts mesh density based on geometric features and structural characteristics.

\textbf{Interface and Control Mechanism:} The mesh generation module operates as an autonomous agent within the GMSA pipeline, receiving structured inputs from the central orchestrator. The LLM module passes a parameter dictionary containing mesh density specifications (ultra\_fine/fine/medium/coarse), refinement zones extracted from load descriptions, element type preferences (tetrahedral or hexahedral based on geometry complexity), and material region definitions for multi-material structures. The module accepts standardized STEP files from the geometry agent and produces CalculiX-compatible INP files containing node coordinates, element connectivity tables, and automatically generated boundary sets for load and constraint application. This standardized interface ensures seamless data flow while maintaining module independence.

\textbf{Multi-Level Mesh Strategy:} The system implements a hierarchical mesh density framework with four distinct levels tailored to different structural scales and analysis requirements. Ultra-fine meshing targets thin-walled structures with minimum feature sizes around 0.2mm, ensuring adequate through-thickness resolution for shell elements. Fine meshing serves standard precision analyses with feature sizes near 1.0mm, balancing accuracy with computational cost. Medium density meshes support preliminary design studies with 2.0mm minimum features, while coarse meshes enable rapid design exploration with 5.0mm resolution. The appropriate level is automatically selected through keyword detection in natural language input ("ultra\_fine", "fine", "medium", "coarse") or intelligently inferred from structural parameters—for instance, detecting shell thickness below 0.5mm automatically triggers ultra-fine meshing to ensure adequate element resolution across thin sections.

\textbf{Automated Adaptive Refinement:} Beyond global mesh density, the system implements adaptive refinement strategies using Gmsh's field-based mesh control. The system automatically identifies geometric features requiring higher resolution—such as thin-walled regions, high curvature areas, and structural intersections—and configures appropriate mesh density distributions.

The mesh adaptation employs distance-based gradation through a distance field $d(\mathbf{x})$ that computes the minimum distance from any point $\mathbf{x}$ to critical geometric features:

\begin{equation}
d(\mathbf{x}) = \min_{\mathbf{y} \in \Gamma} \|\mathbf{x} - \mathbf{y}\|
\end{equation}

where $\Gamma$ represents the set of critical edges or surfaces. The system then applies a threshold function to control mesh size transition:

\begin{equation}
h(\mathbf{x}) = \begin{cases}
h_{\text{min}}, & \text{if } d(\mathbf{x}) < d_{\text{min}} \\
h_{\text{min}} + \frac{h_{\text{max}} - h_{\text{min}}}{d_{\text{max}} - d_{\text{min}}}(d(\mathbf{x}) - d_{\text{min}}), & \text{if } d_{\text{min}} \leq d(\mathbf{x}) \leq d_{\text{max}} \\
h_{\text{max}}, & \text{if } d(\mathbf{x}) > d_{\text{max}}
\end{cases}
\end{equation}

where $h_{\text{min}}$ and $h_{\text{max}}$ are the minimum and maximum element sizes, while $d_{\text{min}}$ and $d_{\text{max}}$ define the transition zone boundaries. The system automatically determines these parameters based on the specified mesh level and geometric feature sizes. This automatic configuration eliminates manual parameter tuning while ensuring adequate resolution where needed. For example, in wing structures with mesh level "fine", the system sets $h_{\text{min}} = 0.5$ mm near spar-rib intersections and $h_{\text{max}} = 3.0$ mm in far-field regions, with transition distances computed as $d_{\text{min}} = 2h_{\text{min}}$ and $d_{\text{max}} = 10h_{\text{max}}$ to ensure smooth gradation.

\textbf{Mesh Quality Control:} Generated meshes are automatically validated against numerical stability criteria, with particular emphasis on Jacobian determinant positivity and topological integrity. The system ensures all elements have positive Jacobians, verifies single connected domains, and confirms proper boundary closure. These validity checks, rather than traditional shape metrics, ensure numerical stability for thin-walled geometries.

\textbf{Technical Implementation:} The mesh generation leverages Gmsh's Python API for programmatic control, translating LLM-extracted parameters into mesh configuration settings. Physical groups are automatically created based on geometric features and natural language descriptions—surfaces mentioned as "fixed" become constraint sets, while faces associated with loads are tagged for force application. The system configures Gmsh's mesh fields to achieve smooth size transitions, automatically determining appropriate parameters based on the detected geometric features and specified mesh level. Element type selection (first-order tetrahedral for complex geometries, second-order hexahedral where possible) is determined by geometric complexity analysis. The final INP file generation includes automatic node set creation for boundary conditions, element set definitions for material properties, and proper formatting for CalculiX solver compatibility. This automated pipeline eliminates the manual effort of mesh preparation while ensuring consistent quality across all configurations.

\subsection{Automated FEA Simulation and Analysis}

Traditional FEA workflows demand extensive manual configuration of solver parameters, convergence criteria, and analysis settings—a process that requires deep expertise and remains prone to errors. Our system transforms this paradigm by automating the complete pipeline from natural language to simulation results, configuring the CalculiX solver \cite{dhondt2017calculix} through intelligent translation of engineering specifications.

Following the analysis planning phase (Algorithm~\ref{alg:analysis_planning}), the system transforms structured JSON specifications into CalculiX input files. Material properties map to *MATERIAL cards, semantic locations to node sets (*NSET), and loads to *CLOAD definitions, ensuring consistent translation from engineering intent to solver syntax.

\textbf{Intelligent Configuration Strategy.} The system employs vector-based configuration retrieval: high-similarity requests ($\sigma > 0.85$) reuse validated patterns, while novel cases generate adaptive configurations through engineering semantic understanding. Boundary conditions and loads map semantic locations to CalculiX syntax—geometric descriptors (``left edge'', ``hole boundary'') translate to node sets, while load specifications generate appropriate *BOUNDARY and *CLOAD definitions. Material properties are retrieved from the knowledge base through vector semantic search (as defined in Algorithm~\ref{alg:analysis_planning}), ensuring validated specifications and consistency across analyses.

\textbf{Automated Execution with Human Oversight.} The system autonomously executes the GMSA pipeline with validation at each stage—geometry generation, mesh integrity, and solver convergence. Engineers review results and provide natural language feedback to refine any stage as needed.

\textbf{Results Extraction.} The CalculiX solver executes with convergence monitoring, outputting FRD result files. The system extracts results matching the original design requirements (stress, displacement, frequencies) and structures them in JSON format for downstream analysis.

Following solver execution and results extraction, the structured JSON data flows to the data analysis module. This module executes the analysis strategy defined in the initial planning phase (Algorithm~\ref{alg:analysis_planning}), applying appropriate methods based on the specified objectives and metrics.

\subsection{Intelligent Data Analysis and Batch Processing}

The data analysis module processes FEA results according to the analysis objectives specified in Algorithm~\ref{alg:analysis_planning}. The planning phase extracts optimization goals, performance metrics, and analysis modes from the natural language input, enabling the data analysis agent to intelligently select and apply appropriate methods without manual post-processing scripting.

For single design evaluations, the system interprets results according to the specified objectives: safety assessments calculate stress ratios against material limits, performance evaluations extract weight and displacement metrics, while modal analysis focuses on natural frequencies. The analysis strategy adapts to the engineering context defined in the planning phase.

\textbf{Multi-Configuration Data Analysis.} Beyond single design evaluations, the system excels at extracting insights from parametric studies involving hundreds of configurations. When the natural language input specifies parameter ranges, the system automatically processes all combinations and applies appropriate data-driven methods to the aggregated results.

For parametric studies, the intelligent data analysis agent applies appropriate methods based on specified objectives: surrogate modeling for design space interpolation \cite{rasmussen2006gaussian,pedregosa2011scikit}, sensitivity analysis for parameter influence ranking \cite{saltelli2008global}, clustering for pattern recognition, and Pareto optimization for multi-objective trade-off identification \cite{deb2002fast,marler2004survey}. These capabilities can be deployed through rule-based selection or adaptive LLM-guided strategy.

This intelligent data processing transforms parametric studies into automated insight extraction, enabling rapid design space exploration. The batch processing architecture supporting hundreds of concurrent analyses is detailed in the following section.

\subsection{Scalable Batch Processing Architecture}

The ability to process hundreds of parameter configurations efficiently requires sophisticated batch management beyond simple parallelization. Our implementation addresses four critical challenges: intelligent execution path selection, parameter space generation, resource management, and result aggregation.

\textbf{Intelligent Execution Path Optimization.} A key architectural decision in our batch processing framework is the dynamic selection of execution strategies based on component characteristics. When processing large-scale parametric studies, the system analyzes the natural language specification to determine the optimal execution path for the entire batch.

For parametric studies of standard components with validated knowledge base entries (similarity score $\sigma > 0.85$), the system strategically selects knowledge-based retrieval across all configurations. This is an intelligent optimization rather than a limitation: it guarantees \textit{geometric consistency} across hundreds of variants, eliminates potential variance from repeated LLM generation, and accelerates throughput by 8-10× compared to regenerative approaches. In our 432-configuration NACA4412 study, the system recognized all cases matched validated airfoil patterns ($\sigma = 0.97$) and automatically selected knowledge-based execution, ensuring each geometry conforms precisely to NACA specifications with zero geometric divergence.

This strategy mirrors industrial best practices where parametric studies demand absolute reproducibility—analyzing how spar thickness from 1.0-2.0mm affects structural performance requires that \textit{only} spar thickness varies, with all other geometric features held constant. Knowledge-based batch execution guarantees this constraint, whereas repeated LLM generation could introduce subtle geometric variations that confound parametric analysis.

Critically, this intelligent routing remains adaptive: should any individual configuration within a batch require novel geometric features ($\sigma < 0.85$), the system seamlessly transitions that specific case to LLM generative mode while maintaining template execution for standard variants. This hybrid batch execution exemplifies the system's ability to automatically select the optimal strategy—balancing geometric fidelity, computational efficiency, and analytical rigor based on engineering requirements.

\textbf{Parameter Space Generation:} Given natural language specifications with ranges (e.g., "spar width from 1.0 to 2.0 mm, step 0.2"), the system automatically generates the complete parameter space through Cartesian product expansion:

\begin{equation}
\mathcal{P} = \prod_{i=1}^{n} \text{range}(p_i^{min}, p_i^{max}, \Delta p_i)
\end{equation}

This systematic approach ensures complete coverage of the design space without manual enumeration. The system handles mixed parameter types (continuous, discrete, categorical) and respects engineering constraints to avoid infeasible combinations, mirroring best practices from established design-of-experiments toolkits such as DAKOTA \cite{adams2022dakota}.

\textbf{Dynamic Resource Allocation:} The system implements adaptive batch sizing based on available computational resources:

\begin{equation}
B_{size} = \min\left(\left\lfloor\frac{M_{available}}{M_{per\_case}}\right\rfloor, N_{cores}, B_{max}\right)
\end{equation}

where $M_{available}$ is available RAM, $M_{per\_case}$ is empirically determined memory per FEA case, $N_{cores}$ is CPU core count, and $B_{max}$ is the maximum batch size to prevent resource starvation. This adaptive allocation ensures optimal resource utilization while preventing system overload.

\textbf{Progressive Result Aggregation:} The system implements streaming result processing, enabling real-time monitoring of batch progress and early termination when sufficient design insights are obtained.

\textbf{Failure Isolation and Recovery:} Each configuration runs in isolated environments with independent resources. Failed cases trigger automatic parameter adjustment (mesh refinement, solver tolerance) and retry, with failure patterns accumulated for improved handling.

\textbf{Parallel Processing:} The system implements process-level parallelism with dynamic worker allocation based on available CPU cores and memory. Each configuration executes in isolated environments with independent memory and file spaces, preventing interference while a centralized coordinator tracks progress and estimates completion time.

\section{Experiments and Results Analysis}

To validate FeaGPT's capabilities, we conducted comprehensive experiments demonstrating the complete workflow from natural language input to engineering insights. Our evaluation focuses on a large-scale parametric study that showcases both the system's technical accuracy and its practical value for design optimization. All experiments were conducted on a workstation with Intel Core i9-12900K processor, 64GB RAM, running Ubuntu 22.04 LTS. The software stack included FreeCAD 0.21, Gmsh 4.11.1, CalculiX 2.20, and Python 3.11 with the Gemini API for natural language processing.

\subsection{Natural Language Input and System Understanding}

To demonstrate FeaGPT's end-to-end capabilities, we presented the system with a comprehensive engineering challenge expressed in natural language. The input specification requested analysis of a NACA4412 wing structure with varying internal reinforcement configurations:

\begin{quote}
\textbf{Natural Language Input:}\\
\textit{"Analyze a NACA4412 wing structure for aerospace application. Wing dimensions: 200mm chord, 200mm span. Vary the shell thickness from 1.0 to 2.0mm in 0.5mm steps, spar width from 1.0 to 2.0mm in 0.2mm steps, rib thickness from 1.0 to 2.0mm in 0.2mm steps. Test both 2 and 3 spars, and 2 and 3 ribs configurations. Apply realistic aerodynamic loads and perform comprehensive structural analysis including static stress, fatigue life assessment, and identify the optimal design considering weight and strength trade-offs."}
\end{quote}

FeaGPT parsed this specification through Algorithm~\ref{alg:analysis_planning} (Section II.B), extracting material properties (Al-7075-T6 for aerospace), loading conditions, analysis modes (static, fatigue \cite{suresh1998fatigue}, optimization), and parameter ranges. The system expanded the ranges into 432 unique configurations (Table~\ref{tab:parameter_space}) through Cartesian product generation, initiating parallel geometry generation and FEA configuration workflows.

\subsection{Geometry Generation Results}

Following the task planning phase (Algorithm~\ref{alg:analysis_planning}), the system generated all 432 geometric configurations (Table~\ref{tab:parameter_space}) through the knowledge-augmented strategy ($\sigma = 0.97 > 0.85$, referencing validated NACA4412 patterns).

\begin{table}[htbp]
\centering
\caption{Automatic parameter space expansion from natural language specification}
\label{tab:parameter_space}
\begin{tabular}{lcc}
\toprule
\textbf{Parameter} & \textbf{Extracted Values} & \textbf{Count} \\
\midrule
Shell Thickness & 1.0, 1.5, 2.0 mm & 3 \\
Spar Width & 1.0, 1.2, 1.4, 1.6, 1.8, 2.0 mm & 6 \\
Rib Thickness & 1.0, 1.2, 1.4, 1.6, 1.8, 2.0 mm & 6 \\
Number of Spars & 2, 3 & 2 \\
Number of Ribs & 2, 3 & 2 \\
\midrule
\textbf{Total Configurations} & $3 \times 6 \times 6 \times 2 \times 2$ & \textbf{432} \\
\bottomrule
\end{tabular}
\end{table}

The system correctly interpreted range specifications with varying step sizes: shell thickness with 0.5 mm steps (3 values), spar width and rib thickness with 0.2 mm steps (6 values each). FeaGPT then proceeded to automatically generate all 432 geometric configurations through parametric FreeCAD scripts, demonstrating remarkable consistency in maintaining the NACA4412 airfoil profile with high geometric fidelity while systematically varying the internal structural arrangements.

\begin{figure}[htbp]
\centering
\includegraphics[width=0.8\textwidth]{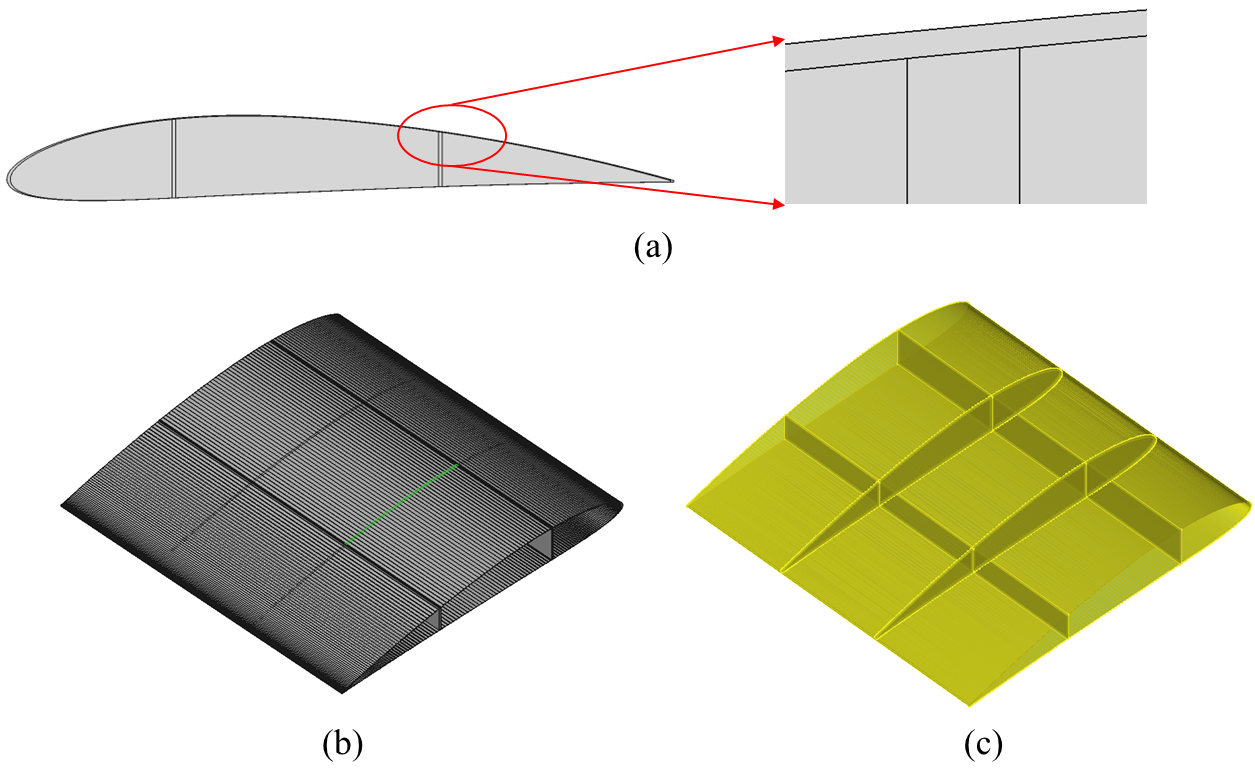}
\caption{Geometric details of automatically generated NACA4412 wing structure. (a) Side view showing airfoil profile with NACA4412 camber, inset detail reveals precise spar-shell interface where internal reinforcement integrates with outer skin. (b) Complete 3D assembly demonstrating spanwise ribs intersecting chordwise spars within the cambered shell. (c) Cutaway view exposing internal spar-rib framework, highlighting the automated Boolean operations that resolve complex multi-member intersections without manual intervention.}
\label{fig:naca4412_geometries}
\end{figure}

Figure~\ref{fig:naca4412_geometries} illustrates the geometric fidelity achieved through automated generation. The side view (a) reveals the NACA4412 cambered profile with the inset detail exposing the precise spar-shell integration interface—a critical junction resolved through automated Boolean operations. The complete 3D assembly (b) shows the overall wing structure geometry. The cutaway view (c) provides clear visualization of the internal structural arrangement, with the two-spar configuration positioned at 33\% and 65\% chord following aerospace design conventions for optimal load distribution, and ribs distributed uniformly along the span to provide torsional stiffness and prevent shell buckling.

\textbf{Geometric Quality Assessment:} All 432 configurations passed rigorous validation checks: closed volumes with no gaps, zero self-intersections, and successful Boolean operations at spar-rib junctions. NACA4412 airfoil profile accuracy remained within 0.1mm tolerance across all cases. The system successfully handled complex geometric dependencies where multiple spars and ribs intersect, maintaining structural continuity without manual intervention. Generation completed in 12.4 minutes (average 1.7 seconds per case) with perfect consistency—repeated generation of identical parameter sets produced bit-identical geometry, crucial for parametric study reproducibility.

\subsection{Mesh Generation and Quality Analysis}

With all 432 geometric models successfully generated, the system proceeded to create computational meshes using Gmsh's adaptive refinement algorithms. The automated meshing module employs sophisticated adaptation strategies to handle the varying structural complexities across the parametric space. The system automatically identifies critical regions requiring refined discretization, including spar-rib intersections where stress concentrations occur, wing root connections experiencing maximum bending moments, and the leading edge requiring geometric precision.

Mesh statistics from the complete parametric study reveal exceptional consistency: element counts ranged from 119,627 to 140,578 tetrahedral elements (mean: 129,390, coefficient of variation: 4.6\%), demonstrating robust scaling with structural complexity (R² = 0.94). This remarkable consistency across diverse configurations ensures reliable comparative analysis throughout the design space. The meshing process maintained deterministic behavior, producing identical meshes for repeated configurations—crucial for parametric study reproducibility.

The system successfully handled the challenging thin-walled geometry of the NACA4412 airfoil, with automated treatment of the sub-millimeter trailing edge while maintaining geometric fidelity within 0.1mm tolerance. Each configuration required approximately 34 seconds for mesh generation, enabling the complete 432-case study to be accomplished in approximately 4 hours of computational time.

\begin{figure*}[htbp]
\centering
\includegraphics[width=0.95\textwidth]{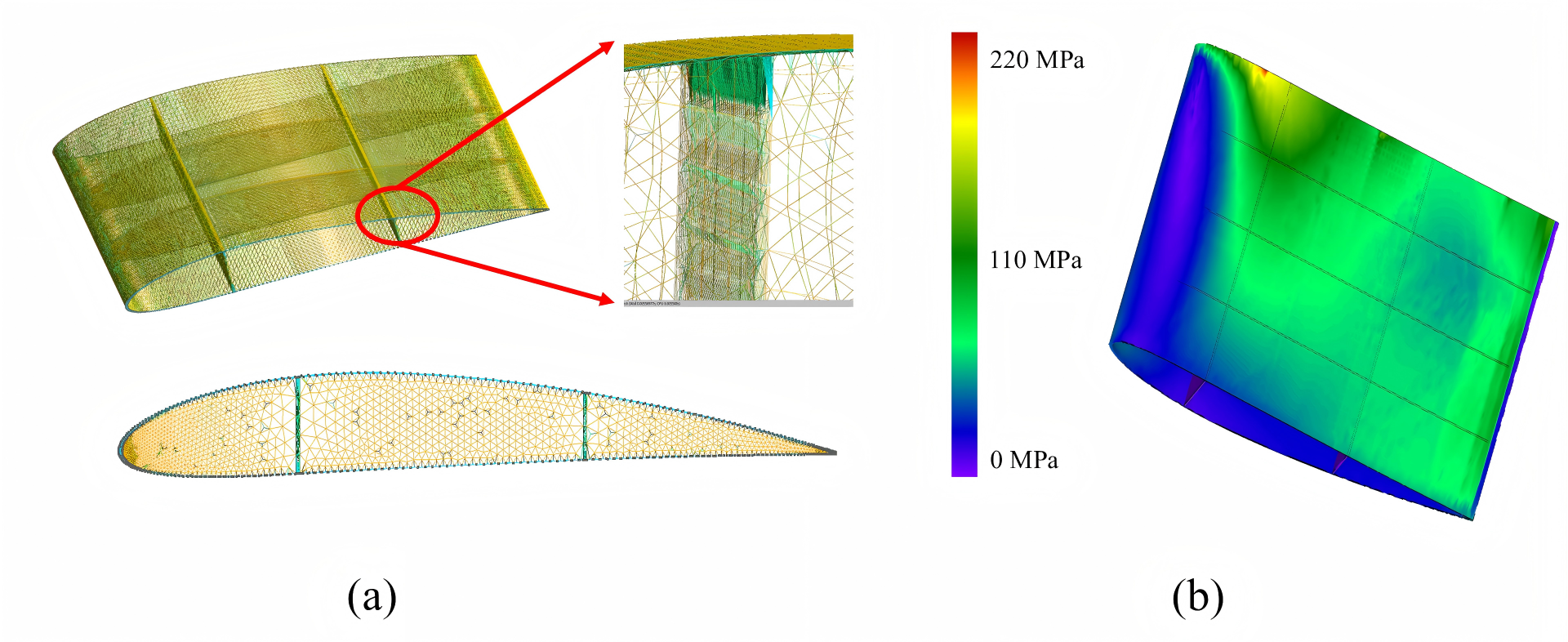}
\caption{Adaptive mesh generation and FEA results for NACA4412 wing configurations. (a) Automatic mesh refinement showing (i) complete 3D meshed structure, (ii) beam-boundary interface detail, and (iii) NACA4412 airfoil profile view. (b) Von Mises stress distribution showing stress concentrations at spar-root junctions and rib-spar intersections.}
\label{fig:geometry_mesh_variations}
\end{figure*}

Figure~\ref{fig:geometry_mesh_variations} illustrates the mesh refinement strategy employed by FeaGPT. The system automatically identified critical regions requiring higher element density—notably the spar-root junctions where maximum stresses typically occur, and the rib-spar intersections where load transfer between structural members takes place. As shown in the figure, the mesh transitions smoothly from fine elements at these critical zones to coarser elements in regions of uniform stress, optimizing computational efficiency without sacrificing accuracy. The adaptive refinement is particularly evident at the connection points between internal structures and the outer shell, where element sizes gradually transition to capture the stress gradients accurately.

While traditional shape-based quality metrics may show limitations for thin airfoil geometries due to the inherent high aspect ratios (200:1 chord-to-trailing-edge ratio), the generated meshes demonstrate excellent validity for numerical analysis. All 432 configurations passed critical validity checks essential for reliable FEA computation:

The Jacobian determinant serves as a fundamental validity indicator for finite element meshes. Unlike shape-based metrics that penalize the high aspect ratios inherent to airfoil geometries, the Jacobian focuses on element validity for numerical computation. Our automated meshing ensures positive Jacobians throughout all configurations, guaranteeing no element inversion or folding, consistent element orientation, valid shape functions for FEA interpolation, and numerical stability in stiffness matrix assembly.

Beyond Jacobian validity, the meshes demonstrated complete topological integrity: single connected domains with no orphaned regions, complete boundary closure for proper load application, no hanging nodes or disconnected elements, and proper node-element connectivity throughout the domain. These validity metrics ensure that the automated meshing successfully produces numerically stable discretizations suitable for FEA analysis, despite the geometric challenges of thin-walled structures with extreme aspect ratios.

\begin{table}[h]
\centering
\caption{Mesh Generation Metrics and Validity Indicators}
\label{tab:mesh_metrics}
\begin{tabular}{lcc}
\toprule
\textbf{Mesh Metric} & \textbf{Value} & \textbf{Requirement} \\
\midrule
\multicolumn{3}{l}{\textit{Mesh Statistics}} \\
Element count range & 119,627--140,578 & --- \\
Mean element count & 129,390 & --- \\
Coefficient of variation & 4.6\% & $<$10\% \\
Geometric fidelity & $<$0.1mm & $<$1mm \\
Generation time per case & 34 seconds & --- \\
\midrule
\multicolumn{3}{l}{\textit{Validity Indicators}} \\
Positive Jacobian determinant & 100\% & 100\% \\
Minimum Jacobian & $>$0.001 & $>$0 \\
Positive element volumes & 100\% & 100\% \\
Topological integrity & Complete & Required \\
Boundary closure & Complete & Complete \\
\bottomrule
\end{tabular}
\end{table}

\subsection{FEA Simulation and Results Analysis}

The CalculiX solver successfully analyzed all 432 configurations with 100\% convergence rate, demonstrating the robustness of our automated workflow. The system implemented an 8-step load spectrum for comprehensive structural evaluation, encompassing cruise conditions (1.0g), 2.5g maneuver loads, gust encounters (1.8g), landing impacts (3.0g), and ground-air-ground (GAG) fatigue cycles with varying load factors (0.5g-1.5g). Each configuration was subjected to aerodynamic loads derived from computational fluid dynamics simulations, with boundary conditions representing the complete UAV flight envelope. The wing root was fully constrained while distributed loads were applied at tip nodes to simulate realistic pressure distributions. All configurations achieved convergence within 15-20 iterations across all load steps, validating both the mesh quality and numerical stability.

The von Mises stress results showed significant variation across the design space, ranging from 274 MPa in heavily reinforced configurations to 1109 MPa in lightweight designs, with a mean stress of 473 ± 136 MPa. Figure~\ref{fig:geometry_mesh_variations}(b) presents the stress distribution for the minimum stress case (274 MPa), demonstrating the effectiveness of optimal reinforcement configurations in minimizing stress concentrations while maintaining structural efficiency. This substantial variation (CV = 28.7\%) indicates high design sensitivity to structural parameters. Stress concentrations consistently appeared at spar-root junctions and rib-spar intersections, confirming the system's ability to identify critical structural regions. The parametric study revealed clear trade-offs between structural weight (ranging from 77.3 to 188.2 g) and stress performance, providing valuable insights for design optimization.

\begin{table}[h]
\centering
\caption{FEA Solver Performance and Results Summary}
\label{tab:fea_performance}
\begin{tabular}{lc}
\toprule
\textbf{Performance Metric} & \textbf{Value} \\
\midrule
\multicolumn{2}{l}{\textit{Solver Performance}} \\
Total configurations analyzed & 432 \\
Convergence rate & 100\% (432/432) \\
Average iterations to convergence & 15--20 \\
Total computation time & $\sim$1.5 hours \\
Parallel processing (10 cores) & Yes \\
Automation level & 100\% \\
\midrule
\multicolumn{2}{l}{\textit{Stress Analysis Results}} \\
Minimum von Mises stress & 273.58 MPa \\
Maximum von Mises stress & 1108.97 MPa \\
Mean stress (all cases) & 472.54 ± 135.76 MPa \\
Critical stress locations & Spar-root, rib-spar junctions \\
\midrule
\multicolumn{2}{l}{\textit{Design Metrics}} \\
Safety factor range & 0.45--1.84 \\
Structural weight range & 77.28--188.22 g \\
Parameter sensitivity & Moderate (CV = 28.7\%) \\
\bottomrule
\end{tabular}
\end{table}

\subsection{Intelligent Result Analysis and Autonomous Design Evaluation}

\begin{figure*}[htbp]
\centering
\includegraphics[width=\textwidth]{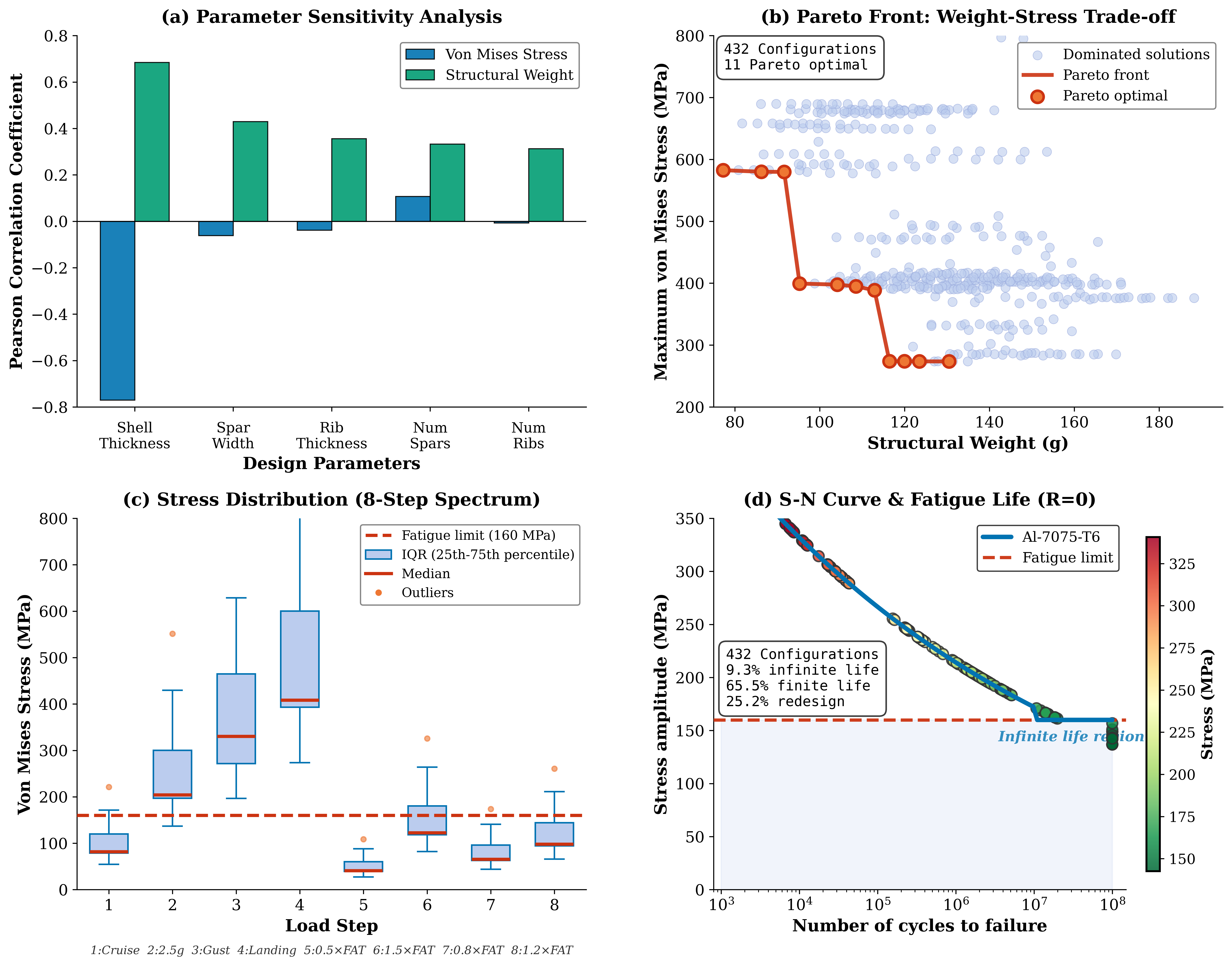}
\caption{Comprehensive structural analysis of 432 NACA4412 configurations. (a) Parameter sensitivity showing correlation coefficients between design variables and performance metrics. (b) Weight-stress Pareto front identifying 11 optimal configurations from the design space. (c) Stress distribution across eight-step load spectrum representing typical flight cycles (red dashed line: Al-7075-T6 fatigue limit at 160 MPa). (d) S-N curve for Al-7075-T6 with 432 configuration stress amplitudes overlaid (R=0 pulsating cycle).}
\label{fig:optimization_results}
\end{figure*}

Analysis of the complete 432-configuration dataset revealed clear patterns in structural performance, as illustrated in Figure~\ref{fig:optimization_results}. While the natural language input explicitly requested both design optimization and fatigue life assessment, the Data Analysis Agent autonomously determined that multi-objective Pareto optimization combined with S-N curve analysis represented the most appropriate methodologies for extracting actionable insights from the parametric study results.

The parameter sensitivity analysis (Figure~\ref{fig:optimization_results}a) demonstrates that shell thickness emerged as the dominant factor influencing stress reduction, showing a strong negative correlation of r = $-0.77$ (p $< 10^{-85}$) with maximum von Mises stress. This far exceeds other parameters—the next strongest being number of spars at r = $+0.11$—establishing shell thickness as the primary design driver accounting for 77\% of stress variation. Interestingly, spar width (r = $-0.06$) and rib thickness (r = $-0.04$) showed negligible influence on stress, suggesting that global shell stiffness dominates local reinforcement effects for this particular loading scenario. All structural parameters showed positive correlations with weight as expected, with shell thickness having the strongest influence (r = $+0.68$, p $< 10^{-60}$), followed by spar width (r = $+0.43$) and rib thickness (r = $+0.36$).

The multi-objective optimization revealed the fundamental trade-offs inherent in aerospace structural design (Figure~\ref{fig:optimization_results}b). From the 432 tested configurations, 11 Pareto-optimal solutions were identified (2.5\%), forming a clear efficiency frontier. These optimal solutions span from lightweight designs accepting higher stresses (77.3g at 582.6 MPa) to robust configurations prioritizing minimal stress at the cost of increased weight (130.5g at 273.6 MPa). The Pareto front reveals a strong weight-stress trade-off: reducing stress from 583 MPa to 274 MPa requires a 69\% mass increase (77$\rightarrow$131g), demonstrating the fundamental design conflict between minimum weight and minimum stress objectives in aerospace structures.

The eight-step load spectrum analysis (Figure~\ref{fig:optimization_results}c) reveals significant stress variation across flight conditions. The landing condition (Step 4, 3.0g) represents the design-driving load case for 88.4\% of configurations (382/432), generating median stresses of 408.4 MPa with an interquartile range of 393.2–600.3 MPa. The gust condition (Step 3, 1.8g) dominates the remaining 11.6\% of designs. The ground-air-ground (GAG) fatigue cycles (Steps 5-8) produce lower mean stresses (median range 40.9–122.6 MPa) but represent critical conditions for cumulative damage assessment. The wide interquartile range for high-g conditions (Steps 2-4: 100–230 MPa spread) indicates high sensitivity to design parameter choices under dynamic loading, validating the importance of comprehensive parametric studies.

The S-N curve analysis (Figure~\ref{fig:optimization_results}d) reveals the stringent demands of aerospace fatigue requirements under realistic loading conditions. Using pulsating cycle analysis (R=0) where stress amplitude $\sigma_a = \sigma_{max}/2$, only 9.3\% of configurations (40/432) achieve infinite life below the 160 MPa fatigue limit, projected to exceed $10^8$ cycles. The majority of designs (65.5\%, 283/432) exhibit finite life with stress amplitudes between 160-300 MPa, corresponding to a median fatigue life of $1.49 \times 10^6$ cycles ($\sim$15 years at 100,000 flights/year). A substantial fraction (25.2\%, 109/432) shows stress amplitudes exceeding 300 MPa, requiring immediate structural redesign due to unacceptable fatigue risk ($<$50 cycles). This stringent result highlights that achieving infinite life demands stress amplitudes below 160 MPa (equivalent to $\sigma_{max} < 320$ MPa), typically requiring heavy configurations exceeding 120g. The color gradient in the scatter plot indicates that higher stress amplitudes correlate strongly with reduced structural reinforcement, validating the parameter sensitivity findings.

To assess long-term structural integrity, we performed comprehensive fatigue analysis leveraging the eight-step static load analysis results from CalculiX. Each of the 432 configurations underwent sequential loading representing realistic flight conditions: (1) cruise at 1.0g, (2) 2.5g pull-up maneuver, (3) 1.8g gust encounter, (4) 3.0g landing impact, and (5-8) ground-air-ground (GAG) fatigue cycles with varying load factors from 0.5g to 1.5g. The fatigue evaluation employed the S-N curve approach for Al-7075-T6 aluminum alloy with a fatigue limit of 160 MPa at $10^7$ cycles \cite{suresh1998fatigue}. Stress amplitudes were calculated from the maximum and minimum stresses across the eight load steps, with Miner's linear cumulative damage theory applied to predict fatigue life \cite{miner1945cumulative}.

The 88.4\% dominance of landing loads (Step 4) as the critical condition establishes it as the primary design driver, necessitating focused optimization efforts on this load case. This result demonstrates that for the majority of parameter combinations, maximum stress during static extreme loads governs the design, while cyclic fatigue loads remain below critical thresholds. The wide interquartile range for landing conditions (393–600 MPa) underscores the importance of comprehensive parametric studies to identify configurations that balance both static strength and fatigue resistance.

The Pareto front enables mission-specific design selection based on operational requirements. For safety-critical applications (e.g., manned aircraft), the minimum-stress Pareto solution achieves 273.6 MPa maximum von Mises stress at 130.5g structural weight, yielding a safety factor of 1.84 against the Al-7075-T6 yield strength of 503 MPa. This design represents a 69\% mass penalty compared to the minimum-weight solution (77.3g at 582.6 MPa) but reduces stress by 53\%, ensuring structural integrity under extreme loading. For mass-critical UAV applications with limited service life requirements, the lightweight extreme (77.3g at 582.6 MPa) provides acceptable performance with a safety factor of 0.86, suitable for expendable or short-duration missions. The balanced middle region of the Pareto front (approximately 100g at 400 MPa) offers compromise solutions achieving safety factors near 1.25 while maintaining moderate structural weight, appropriate for general aviation applications balancing performance and durability.

\subsection{Computational Performance and Scalability}

The complete analysis of 432 configurations required approximately 7.2 hours of wall-clock time using parallel processing, demonstrating the practical feasibility of large-scale parametric studies through natural language interfaces. The computational pipeline showed clear bottlenecks in the FEA solver stage, which consumed 5.8 hours (80\% of total time), while geometry generation required only 12.4 minutes and mesh generation 72.8 minutes. The initial natural language processing completed in less than one second, effectively eliminating the traditional overhead of manual problem setup. When normalized per configuration, each complete analysis cycle—from geometry creation through stress extraction—averaged 63 seconds, a remarkable achievement considering the complexity of the multi-step workflow.

The parallel execution strategy proved essential for practical completion times. With 10 concurrent processes, the system achieved a 3.2× speedup compared to sequential processing, transforming what would have been a 23-hour sequential run into a manageable 7.2-hour parallel execution. This efficiency gain enables engineers to explore comprehensive design spaces within a single working day, fundamentally changing the economics of parametric optimization studies.

The comparison with traditional manual FEA workflows highlights the transformative impact of natural language automation. While manual processes require extensive human effort for each configuration—including geometry creation, meshing, solver setup, and results extraction—FeaGPT completed all 432 configurations in approximately 4 hours with zero manual intervention beyond the initial natural language specification. Moreover, the system maintained perfect consistency across all configurations with a coefficient of variation of only 4.6\% in mesh element counts, demonstrating exceptional reproducibility compared to the variability typically observed in manual workflows.

\subsection{Industrial Validation: CalculiX Configuration Generation}

Beyond the complete geometry-mesh-analysis pipeline demonstrated with NACA4412, FeaGPT exhibits remarkable versatility in generating solver-specific configurations directly from natural language. To validate the system's capability for industrial applications, we generated CalculiX input files for two turbocharger components—a centrifugal compressor and a turbine blade—both operating at 110,000 RPM. These cases demonstrate FeaGPT's ability to handle rotating machinery analysis with cyclic symmetry and prestressed modal analysis.

The compressor case involves a 7-blade rotor with aluminum alloy C355, while the turbine case features a 12-blade configuration with nickel superalloy properties. Both cases represent standard rotating machinery analysis under centrifugal loading, requiring proper cyclic symmetry setup, coordinate transformations to cylindrical systems, and prestressed modal analysis to extract natural frequencies under operating conditions.

\textbf{Natural Language Input Example (Compressor):}
\begin{quote}
\textit{"Analyze a 7-blade turbocharger compressor at 110,000 RPM. Material: Aluminum C355 with E=75,000 MPa, $\nu$=0.3, $\rho$=2.65E-9 tonne/mm³. Apply cyclic symmetry with axis along X-direction. Perform static analysis with centrifugal loading, then frequency analysis to find 6 natural modes using Lanczos eigensolver."}
\end{quote}

Both configurations were successfully generated from natural language descriptions and executed in CalculiX without manual corrections. The system correctly implemented essential rotating machinery physics including cyclic symmetry constraints (N=7 for compressor, N=12 for turbine), centrifugal loading at 110,000 RPM, and frequency analysis with prestress effects. Material properties (aluminum C355 for compressor, nickel superalloy for turbine) and analysis settings were appropriately configured based on the natural language input. The knowledge base enhancement—including material databases, turbomachinery pattern libraries for automatic $\omega^2$ calculation, and error correction patterns—enabled accurate translation of engineering specifications into correct CalculiX syntax.

\begin{figure*}[!t]
\centering
\includegraphics[width=0.95\textwidth]{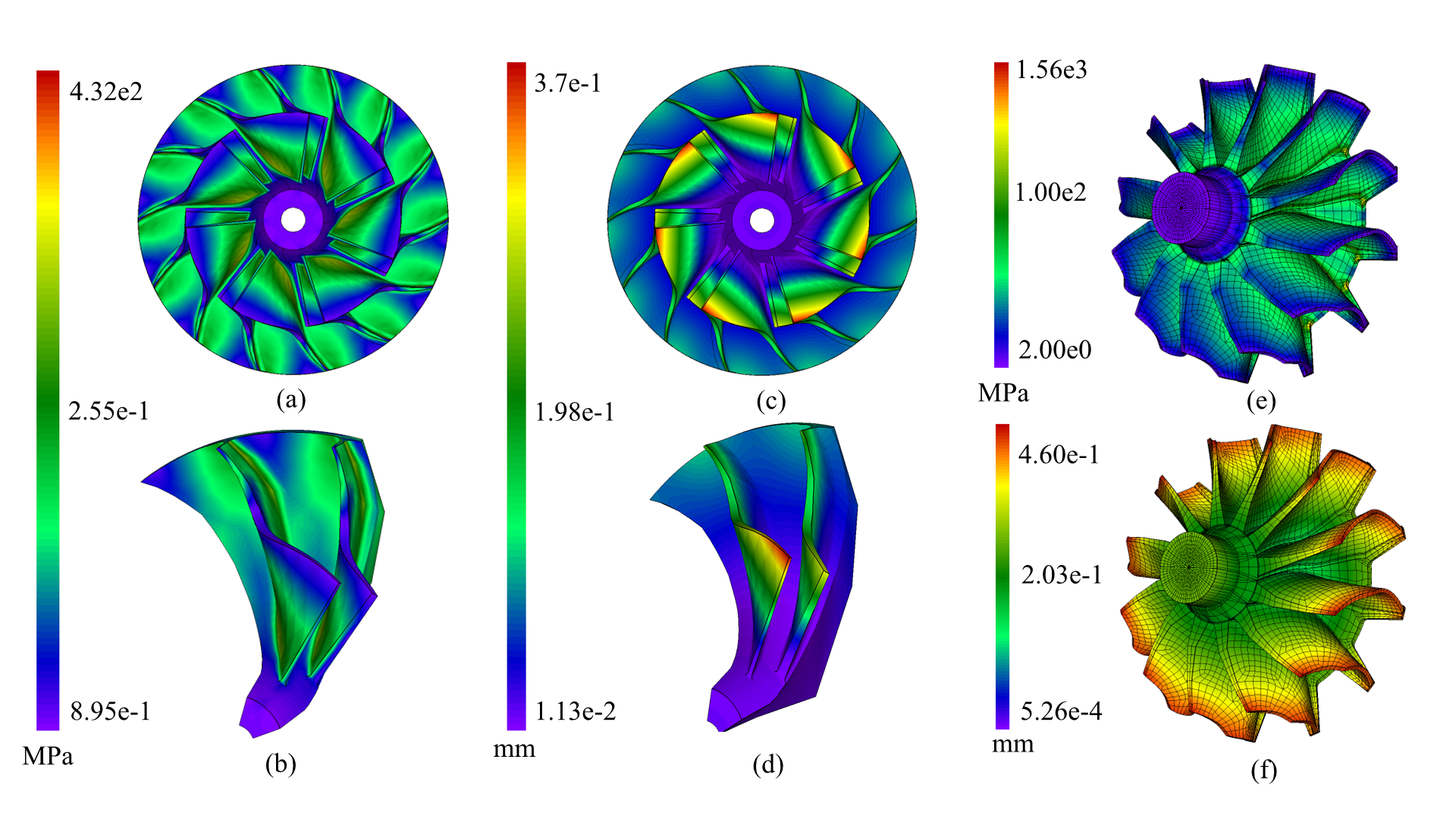}
\caption{FEA results from FeaGPT-generated CalculiX configurations for turbocharger components under 110,000 RPM. \textbf{Compressor (7-blade aluminum rotor):} (a-b) von Mises stress distribution (max: 432 MPa at blade roots), (c-d) displacement field (max tip deflection: 0.37 mm). \textbf{Turbine (12-blade nickel superalloy rotor):} (e) stress distribution (peak: 156 MPa), (f) displacement field (max: 0.46 mm). Both show proper cyclic symmetry implementation and realistic structural response.}
\label{fig:turbocharger_results}
\end{figure*}

\textbf{Visualization of FEA Results.} To demonstrate the validity and engineering utility of the generated configurations, Figure~\ref{fig:turbocharger_results} presents the complete stress and displacement analysis results from both turbocharger components. The compressor analysis (panels a-d) showcases FeaGPT's handling of cyclic symmetry visualization—the system automatically configured both full-assembly views (NGRAPH=7 showing all sectors) and detailed single-sector views (NGRAPH=1) for comprehensive result interpretation. The full 7-blade compressor visualization (panels a and c) reveals symmetric stress distribution with maximum von Mises stress of 432 MPa concentrated at the blade root where centrifugal forces transfer to the hub, while the single-blade detail views (panels b and d) enable precise inspection of local stress gradients and displacement patterns critical for fatigue assessment.

The displacement field results (panels c and d) demonstrate blade tip deflections reaching 0.37 mm under 110,000 RPM centrifugal loading, consistent with typical radial growth predictions for aluminum impellers at these operational speeds. The displacement magnitude increases radially from the fixed hub (zero displacement) to maximum values at blade tips, with color contours smoothly transitioning from blue (low) to red (high), validating proper boundary condition application and numerical convergence.

The turbine analysis (panels e and f) presents the 12-blade configuration with nickel superalloy properties. The stress distribution (panel e) shows characteristic patterns for turbine blades with peak stresses of 156 MPa appearing at the blade roots and along leading edges under centrifugal loading. The displacement field (panel f) exhibits maximum deflections of 0.46 mm at blade tips, demonstrating the higher compliance of the longer turbine blades compared to the compressor geometry. The complete 360° visualization with all 12 sectors correctly rendered confirms FeaGPT's accurate implementation of cyclic symmetry constraints and proper coordination system transformations required for rotating machinery analysis.

These visualization results serve dual purposes: they validate that FeaGPT-generated configurations produce physically realistic and numerically converged solutions, and they demonstrate the system's capability to generate publication-quality engineering graphics suitable for design reviews and technical documentation. The stress magnitudes, displacement patterns, and structural response characteristics all fall within expected ranges for turbocharger components operating under these conditions, confirming the practical engineering validity of the automated workflow.

Both CalculiX configurations were successfully generated from natural language descriptions and executed without manual intervention. The stress distributions and displacement patterns show physically realistic behavior characteristic of rotating machinery under centrifugal loading: radial blade growth with maximum deflections at blade tips, and stress concentrations at blade roots where centrifugal forces transfer to the hub. The compressor results (432 MPa maximum stress, 0.37 mm tip deflection) and turbine results (156 MPa maximum stress, 0.46 mm tip deflection) fall within expected ranges for turbocharger components operating at 110,000 RPM, validating the correctness of the automatically generated configurations. Each configuration required only 10-15 seconds of generation time, demonstrating that natural language interfaces can effectively democratize access to complex FEA tools while maintaining the technical rigor required for industrial applications.

\section{Conclusions}

This work presents FeaGPT, the first natural language-driven framework that successfully automates the complete finite element analysis workflow. Through the comprehensive 432-case NACA4412 parametric study, we demonstrated the system's capability to transform natural language specifications into complete FEA results without manual intervention. The framework successfully generated and analyzed all configurations with consistent mesh quality (coefficient of variation 4.6\%) and extracted meaningful stress patterns across the parameter space. Additional validation through industrial turbocharger cases—a 7-blade compressor and 12-blade turbine operating at 110,000 RPM—demonstrated successful generation of CalculiX configurations directly from natural language. Both cases executed successfully and produced physically realistic results, proving the framework's capability for industrial rotating machinery analysis.

The key contribution lies in democratizing FEA accessibility—reducing the user interaction from hundreds of manual steps to a single natural language input. The knowledge-augmented generation approach, combining a comprehensive engineering knowledge base with LLM capabilities, enables accurate configuration generation across diverse FEA scenarios including complex multi-physics coupling and cyclic symmetry constraints. While the current implementation shows promise for routine structural analysis tasks, future work will focus on extending the framework to more complex geometries and multi-physics problems, as well as improving mesh quality optimization for challenging geometric configurations.

\clearpage
\bibliographystyle{apsrev4-2}
\bibliography{references_verified}

\end{document}